\documentstyle[12pt,epsfig]{article}
\begin{document}
\begin{center}
{\bf ORTHONORMAL POLYNOMIAL APPROXIMATION OF MINERAL WATER DATA
WITH ERRORS IN BOTH VARIABLES}

Nina B. Bogdanova {\footnote { Email:nibogd@inrne.bas.bg}} and
Stefan T. Todorov {\footnote {Email:todorov\_st@yahoo.com}}

{\it INRNE,BAS,72 Tzarigradsko choussee,1784 Sofia, Bulgaria}
\end{center}

{\bf  Abstract}

In this paper we introduce the data from mineral water probe with
errors in both variables. For this case we apply our orthonormal
polynomial expansion(OPEM) method to describe the data in the new
error corridor. It  received the approximating curves and their
derivatives including the errors by weighting approach. The
numerical method and approximation results are presented and
discussed. The special criteria are carried out for orthonormal and
evaluated from it usual expansions. The numerical results are shown
in tables and figures.

 \noindent
\medskip
 {\bf Key words:} orthonormal and usual polynomial
 approximation, weighted  approach,
 contact mineral water angle data

\section {Introduction}
       The water spectra method applies a  drop taken from a  water probe to measure the water's state spectrum. In the special experiment the drop is placed on a
        hostaphan folio- Figure 1\cite{1}. During the whole process of evaporation of the drop,
        one measures at equal time intervals the drop contact angle  with the folio. On X-axis one has the values
        of the contact angles  within  fixed angular intervals  and on Y-axis the frequency of measurements of these
        angles.

       To compare different state spectra one normalizes each spectrum  dividing its Y-values by the number
       of all measurements and thus obtains a probability distribution . One can change the
       function
        $\phi(\theta )$ on the  independent  angle $\theta$  to the function of energy variable  F(E)
          using the following Antonov transformation \cite{2}:

      $$ f(E)= b \phi(\theta)/\sqrt{1-(1+bE)^2},$$
          where  $$ b=I(1+cos(\theta_{0}))/\gamma.$$
Here $I = 5.03.1018 m^{-2}$ is the density of water molecules in the
surface layer, $\gamma$ is the surface tension, $\theta_{0}$ - the
initial contact angle.

The so obtained graph after measurements by method in  \cite{1} is
referred to as energy spectrum $F(E)$ of the probe. $E$ is the
energy of Hydrogen bond of investigated water.

\begin{figure}
\begin{center}
\includegraphics[height=.3\textheight]{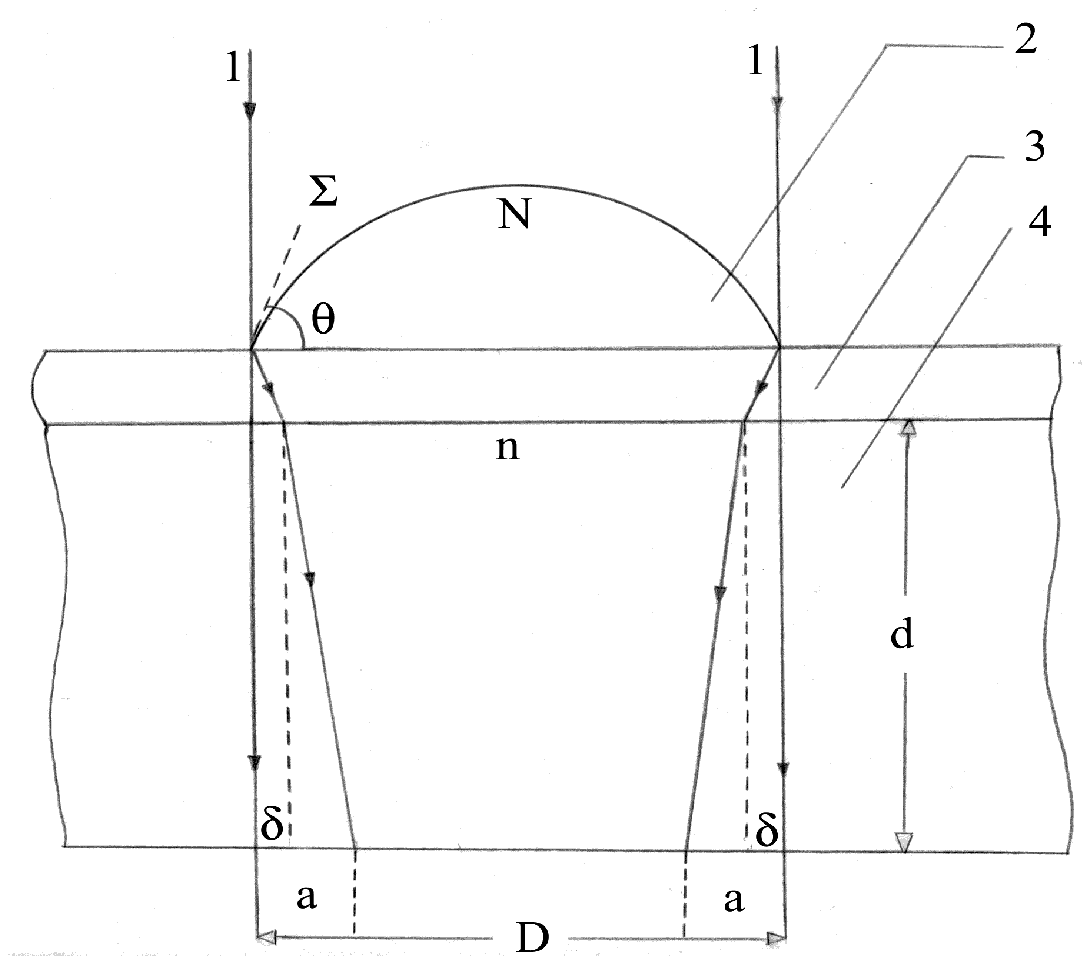}
\caption{\bf Experimental  setup}
\end{center}
\end{figure}


\begin{figure}[!ht]
\begin{center}
\includegraphics[width=13cm]{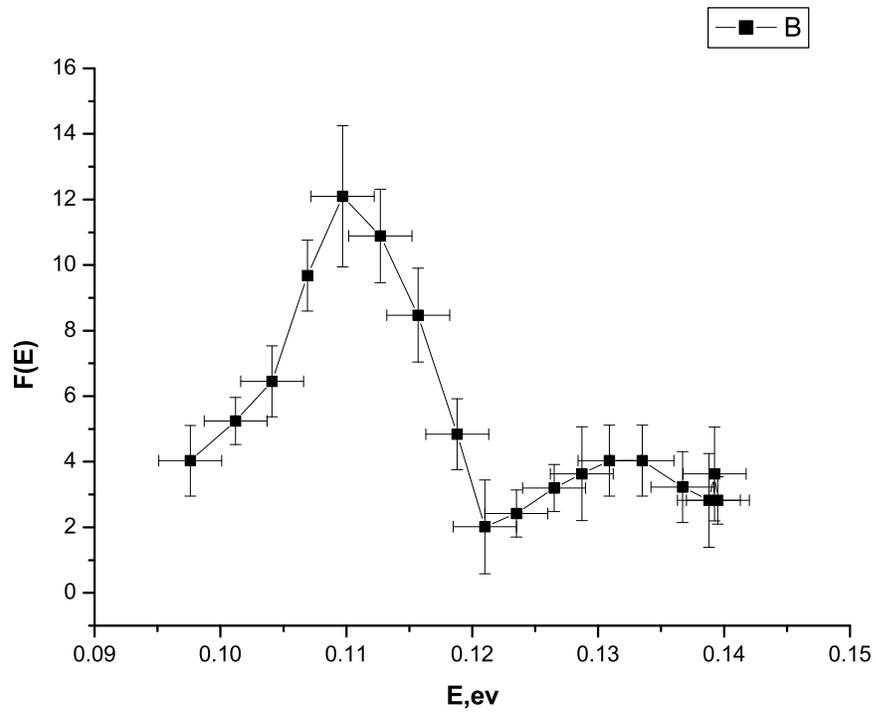}
\end{center}\vspace{-0.4cm}
\caption{\bf Experimental data from  (B) with their
errors}\label{De1}
\end{figure}
On Figure 2 the dependent variable  contains  the values of the
water Hydrogen bond energy.  Here we present new detailed
information about  given data and their errors on both variables
of water probe.

The method of water spectra is sensitive to treatment by  physical
fields as  $\gamma$-ray treatment of water \cite{1,2} and to
environmental changes of the ecosystem on different  water probe
\cite{3,4,5,6,7,8}. In the present paper we approximate  another
natural water data taken from  a water spring in Bulgaria  near
the village Lenovo.

************************

\section {Main  problem definition}

\vspace{4mm}
\begin{itemize}
    \item
To find the best approximation curves  of measured water data on
Fig.2 taking into account the  errors in both variables;
\end{itemize}

\begin{itemize}
    \item
To extend our original Orthonormal polynomial expansion method
(OPEM),  according some criteria, to evaluate orthonormal
description of given data;\end{itemize}

\begin{itemize}
    \item
To find the best approximating  curve with usual polynomials,
evaluated by orthonormal, according some criteria.
\end{itemize}

\section{ Numerical method--OPEM "total variance"}

 Let the
$\{E_{i},F_{i}, i=1,...M \}$ are arbitrary pairs of monitoring
 data $E=E_{i}$ and $F=F_{i}$, introduced in
 section 2. They are given with  experimental errors in both
 variables-$\sigma(F_{i})$ and $\sigma(E_{i})$.
 Consider the total uncertainty (total
 variance) $S^{2}(E,F)$ \cite{9,10,11}, associated with
 $(E,F)$
\begin{equation}
 S_{i}^{2}=\sigma^{2}(F_{i})+({\partial F_{i}\over
\partial E_{i}})^{2}\sigma^{2}(E_{i}),
\end{equation}
according the ideas of Bevington
(1977)\cite{9}, where his proposal is to combine the errors in both
variables and assign them to dependent variable. One defines the
errors corridor $C(E,F)$, which is the set of all intervals
\begin{equation}
[F(E)-S(E,F),F(E)+S(E,F)],
\end{equation}.

\subsection {orthonormal expansion criteria}
 The first criterion to be
satisfied, is that the fitting curve should pass within the errors
corridor $C(E,F)$. In the cases of errors only in $F$, (i.e.
$\sigma(E)=0,\sigma(F)\neq(0))$ the errors corridor $C(E,F)$ reduces
to the  known set of intervals
\begin{equation}
[F-\sigma(F), F+\sigma(F)],
\end{equation}

 for any $F$.
The second criterion is, that the fitting curve $F^{appr}(E_{i})$
satisfies the expression
\begin{equation}
\chi^{2}=\sum_{i=1}^{M}w_{i}[F^{appr}(E_{i})-F(E_{i})]^{2}/(M-L),
w_{i}=1/S_{i}^{2}.
\end{equation}
should be minimal
(L-number of polynomials). The preference is
given to the first criterion. When it is satisfied, the search of
the minimal chi- squared stops.
Some details of the calculation procedure are given in Forsythe's paper \cite{12} and in our works \cite{13,14,15}.\\

Our procedure gives results for approximating function by two
expansions : of orthogonal coefficients $\{a_{i}\}$ and usual ones
$\{c_{i}\}$  with optimal degree $L$:
\begin{equation}
F^{appr(m)}(E)=\sum _{i=0}^{L}a_{i}P_{i}^{(m)}(E)=\sum
_{i=0}^{L}c_{i}E^{i}. \label{3}
\end{equation}
The orthogonal coefficients are evaluated by the given values
$F_{i}$, weights and orthogonal polynomials:
\begin{equation}
a_{i}=\sum _{k=1}^{M}F_{k}w_{k}P_{i}^{(m)}(E_{k}).
\end{equation}
Our recurrence relation for generating orthonormal polynomials and
their derivatives $(m=1,2...)$( or their integrals with
m=-1,-2,-3,...) are carried out by:
\begin{equation}
P_{i+1}^{(m)}(E)= \gamma _{i+1}[E- \mu_{i+1})P_{i}^{(m)}(E)-
(1-\delta _{i0}) \nu _{i}P_{i-1}^{(m)}(E)+m P_{i}^{(m-1)}E)],
\end{equation}
where $\mu_{i}$ and $\nu_{i}$ are recurrence coefficients, and
$\gamma_{i}$ is a normalizing coefficient, defined by scalar
products of given data. One can generate $P_{i}^{m}(E)$
recursively. The polynomials satisfy the following orthogonality
relations: $\sum_{i=1}^{M}w_{i}P_{k}^{(0)}(E_{i})P_{l}^{(0)}
(E_{i})=\delta _{k,l}$ over the discrete point set $\{E_{i},
i=1,2,\ldots\}$. All the calculations for the sake of uniformity
are carried out for $E$ in[-1,1], i.e. after the input interval is
transformed to the unit interval. We remark some advantages of
OPEM: It uses unchanged the coefficients of the lower-order
polynomials; it avoids the procedure of inversion of the
coefficient matrix to obtain the solution, the minimal chi-squared
stops. All these features shorten the computing time and assure
the optimal solution by the criteria(2) and (4).

\subsection{ the usual expansion criteria}

 The
inherited errors in usual coefficients are given by the inherited
errors in orthogonal coefficients:
\begin{equation}
\Delta{c_{i}}=(\sum^{L}_{k=1}(c_{i}^{(k)})^{2})^{1/2}\Delta{a_{i}},
\end{equation}
\begin{equation} \Delta{a_{i}}=
[\sum_{k=1}^{M}P_{i}^{2}(E_{k})w_{k}(F_{k}-F_{k}^{appr})^{2}]^{1/2}.
\end{equation}
where coefficients $c^{(k)}_{i}$ are defined by orthonormal
expansion of polynomials
\begin{equation}
P_{k}=\sum^{k}_{i=0} c^{(k)}_{i}E^{i}, k=0,..,L
\end{equation}
and explicitly constructed by recurrence relation in \cite{13}.

 The procedure is iterative because of the evaluation of
derivatives on every iteration step and the result of the
$k^{it}$-th consequent iteration is called below the $k^{it}$-the
approximation. We note that in every iteration step the algorithm
find the best approximation using given before criteria.

We can add the other criteria for optimal number of polynomials for
usual expansion. Having the $L_{a}$  we continue with finding the
optimal $L_{c}$ the \textbf{minimal value} in
\begin{equation}
 max(
c_{i}(L), i=1,L)
\end{equation}
 in usual coefficients through all steps of
iterations ${k^{it}=1,2,..9}$ or we are asking  the
\textbf{minimal value} of the maximal distance between functions,
evaluated by orthonormal and usual expansions
\begin{equation}
max\vert{(F^{appr}_{a,k}-F^{appr}_{c,k})\vert,k=1,M}
\end{equation}
through all iterations. We investigate both criteria, but we
prefer the last one.


\section { Approximation results}

The main important results from approximation between $2\div 10$
degrees for iterations $1\div 9$  are presented in Table 1 for
characteristics: number of iterations, number of
polynomials,$\chi^{2}$, and $max\vert(F_{a}-F_{c})\vert $. We see
from the Table 1, that from iteration number $2 \div 5$ with
optimal number $L_{a}=6$ the results are good for both expansions
and
 for usual expansion the 8-th
iteration with  optimal number $L_{c}=8$ they are also good.

Note: It is very interesting to present on figures the three curves
- given(B), approximated by orthogonal polynomials(C)  and received
from it by usual polynomials(D)  at different iteration steps.
\begin{table}
 \caption{\label{tab1}\small \bf OPEM approximations results for
every step approximation}
\begin{center}
\begin{tabular}{llllllllll}
\hline
 $k^{it}$ & 1   & 2&   3 & 4 &  5& 6 &7& 8& 9\\
 \hline
  $ L (2 \div 10)$& 7&6 &6&6 &6 &5&6&5&6 \\
  $\chi^{2}*10^{-1}$&
5.61 & 4.23 & 3.99 & 3.79 & 3.77 & 6.81&3.75&6.65&3.63\\
$ max\vert(F
_{a}-F_{c})\vert$&
14.96&3.48&6.75&4.8&4.63&7.53&4.91&0.081&9.33\\
 \hline
 \hline
\end{tabular}
\end{center}
\label{tab:a}
\end{table}

Below the figures 3,4,5,6 present the different approximations
results with 2-nd, 3-rd and 4-th iterations.

\begin{figure}
\begin{center}
\includegraphics[height=.35\textheight]{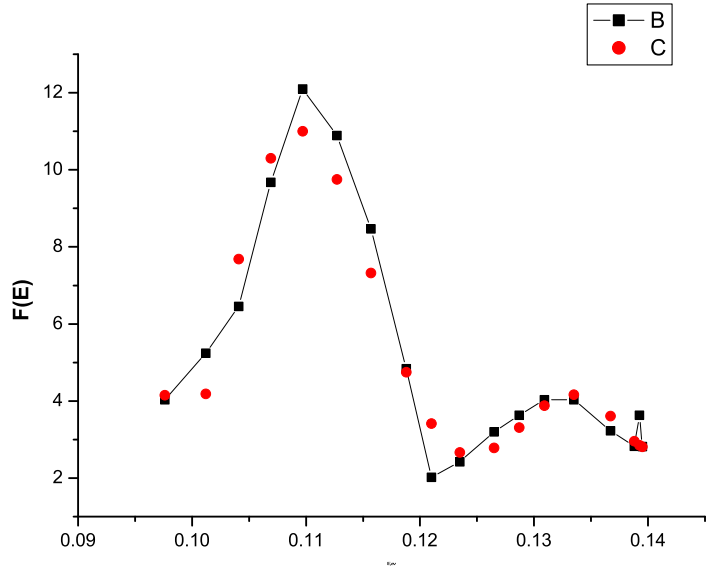}
\end{center}
\caption{\bf OPEM approximation by 6-th degree orthonormal
polynomials (C)(2-nd iteration) of experimental water data(B)}
\label{fig33.eps}
\end{figure}

\begin{figure}
\begin{center}
\includegraphics[height=.35\textheight]{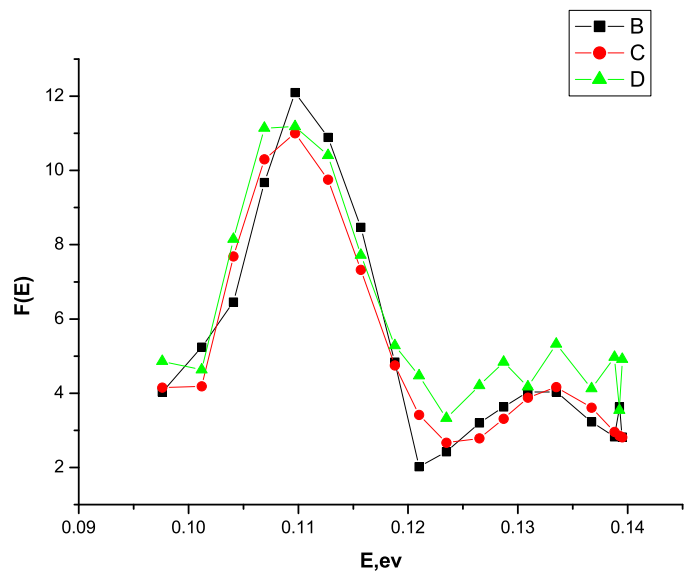}
\end{center}
\caption{\bf OPEM approximation by 6-th degree(iteration-2-nd)
orthonormal polynomials (C) and received usual expansion(D) of
experimental water data(B)} \label{fig55.eps}
\end{figure}

\begin{figure}
\begin{center}
\includegraphics[height=.35\textheight]{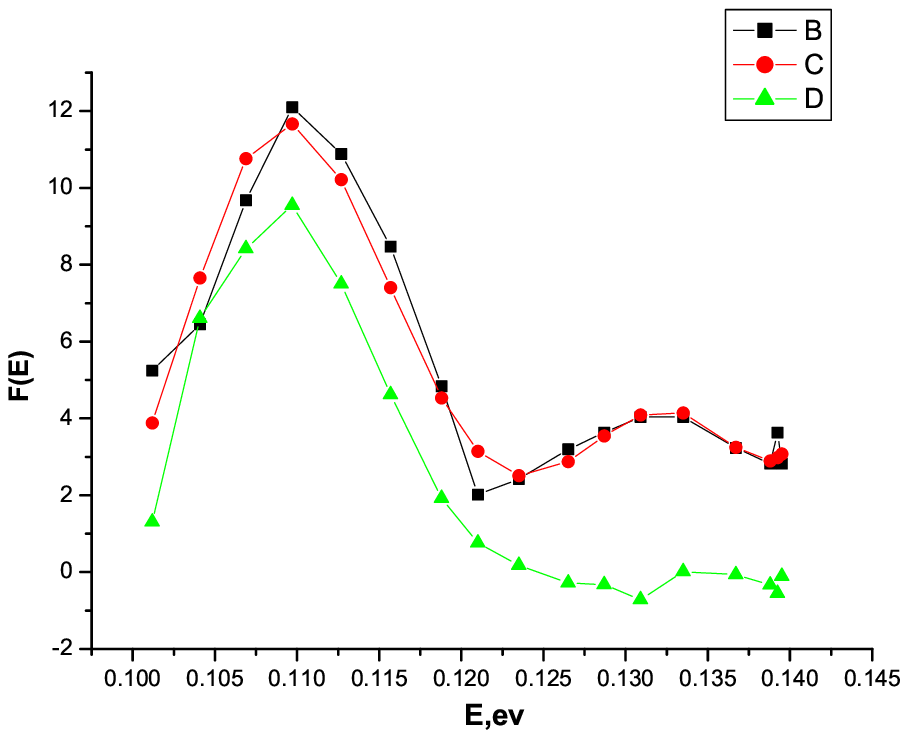}
\end{center}
\caption{\bf OPEM approximation by 6-th degree(the 3-rd
iteration)-the orthonormal polynomials (C) and received usual
expansion(D) of experimental water data(B)} \label{fig66.eps}
\end{figure}


\begin{figure}
\begin{center}
\includegraphics[height=.35\textheight]{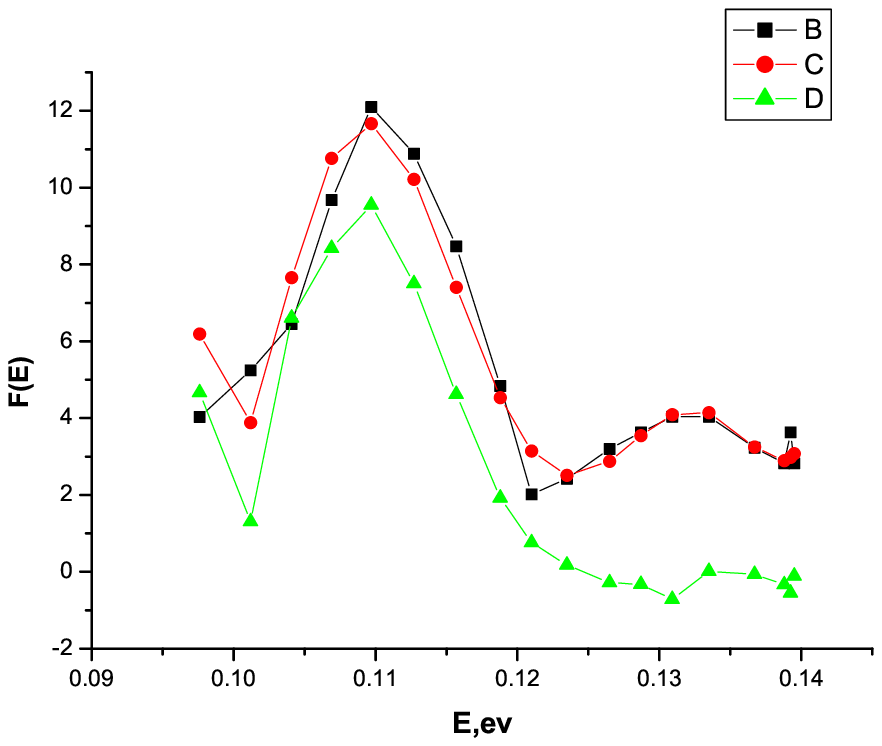}
\end{center}
\caption{\bf OPEM approximation by 6-th degree(the 4-th iteration)
orthonormal polynomials (C) and received usual expansion(D) of
experimental water data(B)} \label{fig77.eps}
\end{figure}

\begin{figure}
\begin{center}
\includegraphics[height=.35\textheight]{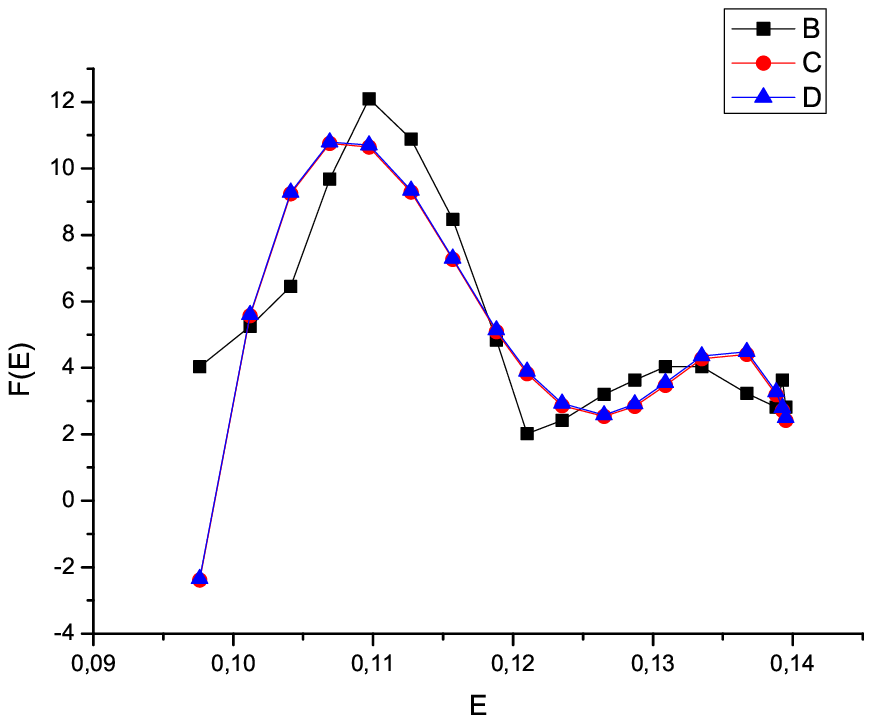}
\end{center}
\caption{\bf OPEM approximation by 5-th degree(the 8-th iteration)
orthonormal polynomials (C) and received usual expansion(D) of
experimental water data(B)} \label{fig88.eps}
\end{figure}

The Table 2 presents the given and approximating values by OPEM
with usual and orthonormal coefficients by calculated optimal
degree $5-th$ in $8-th$ iteration of $M=18$ given values of
following characteristics: energy $E$ , distribution $F$ ,
$\sigma_{E}$ and $\sigma_{F}$, and from $5-th$ column - the
approximating values with orthonormal coefficients
$F^{appr,5}_{a}$, approximating values with usual coefficients
$F^{appr,5}_{c}$, differences $\Delta (F_{a},F_{c})=(
F^{appr,5}_{a}-F^{appr,5}_{c})$, total variance $S(5)$ (equation
(1). The Table  2 shows good coincidence between two descriptions.
 For comparison we can see the previous results for OPEM
applications in \cite {13,14,15,16,17}.

\label{tab2}\label{tab2}
\begin{table}
\caption{ \label{tab2}OPEM approximation of contact water  energy
data}
\begin{tabular}{llrrrrrrr}
\hline
 $ No.$ &$ E[ev] $& $ F(E) $&$\sigma_{E}$& $\sigma_{F}$
&$F^{appr,5}_{a}$&
$F^{appr,5}_{c}$& $  \Delta (F_{a},F_{c})$&S\\
 \hline
1 &0.1395   &  2.820   & 0.025  & 0.72  &2.421            & 2.503   & 8.169-02 & 2.2072 \\
2&0.1392   &  3.627  & 0.025  & 1.43 & 2.721           & 2.799   & 7.796-02 & 2.9469 \\
3&0.1388   & 2.822  & 0.025   & 1.43 & 3.192            & 3.266    & 7.420-02& 2.2173 \\
4&0.1367   & 3.227  & 0.025   & 1.08  & 4.408            & 4.484   & 7.614-02& 1.8114  \\
5&0.1335   &  4.035 &  0.025  & 1.08 &  4.272          &  4.353    & 8.125-02& 1.3297\\
6&0.1309   & 4.035 & 0.025    & 1.08 & 3.467           & 3.549   & 8.161-02 &  1.3126 \\
7&0.1287  & 3.632 & 0.025     & 1.43 & 2.840         & 2.905    &  6.474-02& 2.6050 \\
8&0.1265 & 3.200  & 0.025     & 0.72 &  2.534        &  2.583    &  4.910-02 & 0.9395 \\
9&0.1235 &2.422  & 0.025      & 0.72 & 2.861         &   2.932   & 7.089-02 & 0.5500 \\
10&0.1210&  2.017& 0.025        &1.43  & 3.821      &  3.889   & 6.886-02 & 3.4402 \\
11&0.1188& 4.840& 0.025         & 1.08 & 5.091     & 5.137  & 4.575-02& 5.1487\\
12&0.1157 & 8.470 &0.025        & 1.43  & 7.259       & 7.291    & 3.272-02 & 8.2753 \\
13&0.1127&  10.887 &0.025       & 1.43 & 9.290      &   9.334   & 4.365-02& 5.3774 \\
14&0.1097& 12.095 &0.025        & 2.15  &10.647      &   10.700  &  5.320-02 & 4.6238 \\
15&0.1069& 9.677 &0.025         &  1.08 &10.750        & 10.793    & 4.292-02    & 6.4789\\
16&0.1041&  6.452& 0.025        &  1.08 & 9.243         &  9.276      & 3.293-02       & 15.8508            \\
17&0.1012&  5.242& 0.025        &  0.72 & 5.569         &  5.601         & 3.178-02       & 6.0766     \\
18&0.0975 & 4.030 & 0.025       &  1.08& -2.384    &     -2.347      & 3.714-02          &  86.5354                  \\
 \hline
\end{tabular}
\end{table}

\section {Conclusions}
\begin{itemize}
\item  We have developed new version of OPEM algorithm and Fortran 77 package to include errors in both variables
according (2) and (4), defined new "`total variance"' and taking
into account the respective inherited errors (8) and( 9) in
coefficients.
\end{itemize}
\begin{itemize}
\item The approximating curves are chosen at $2-{nd}, 3-rd , 4-th$
approximation step  by optimal degree $L_{a}=6$ and at 8-th
iteration step by optimal degree $L_{c}=5$
 to satisfy the proposed criteria (2),(4) and (11), (12). The results  show that the orthonormal and
usual expansions values are close to given ones  in the whole
interval.
\end{itemize}
\begin{itemize}
\item Our approximating results with optimal degrees of
orthonormal polynomials for contact (wetting) angle found by
orthogonal and usual coefficients show good \textbf{accuracy and
stability}, demonstrated from Figures  and Tables 1,2. We received
suitable descriptions of the energy variations useful for further
investigations.
\end{itemize}

\begin{itemize}
\item The presented extended algorithm  and package OPEM "total
variance" with its accuracy, stability and speed can be used in
other cases of data analysis (as it it shown in our previous
papers with earlier versions - for calibration problems in high
energy physics \cite{18}).
\end{itemize}

\end{document}